\documentclass[twocolumn]{spie}

\usepackage{amsmath,amsfonts,amssymb}
\usepackage{graphicx}
\usepackage[colorlinks=true, allcolors=blue]{hyperref}
\usepackage{siunitx}

\title{An Associated Particle Imaging System for Soil-Carbon Measurements}

\author[a, b]{Mauricio Ayllon Unzueta}
\author[c]{Eoin Brodie}
\author[d]{Craig Brown}
\author[c]{Cristina Castanha}
\author[d]{Charles~Gary}
\author[e]{Caitlin Hicks Pries}
\author[a]{William Larsen}
\author[a]{Bernhard Ludewigt}
\author[a]{Andrew Rosenstrom}
\author[a]{Arun Persaud}

\affil[a]{Accelerator Technology \& Applied Physics, Lawrence Berkeley National Laboratory,\newline 1 Cyclotron Road, Berkeley, CA 94720, US}
\affil[b]{Nuclear Engineering, University of California, Berkeley, CA 94709, US}
\affil[c]{Earth and Environmental Sciences, Lawrence Berkeley National Laboratory,\newline 1 Cyclotron Road, Berkeley, CA 94720, US}
\affil[d]{Adelphi Technology Inc., 2003 E. Bayshore Road, Redwood City, CA 94063, US }
\affil[e]{Department of Biological Sciences, Dartmouth College, Hanover, NH 03755, US}

\authorinfo{Further author information: (Send correspondence to A. Persaud)\\ A. Persaud: E-mail apersaud@lbl.gov}

\begin{document}
\maketitle

\begin{abstract}
We present first results from experimental data showing the capabilities of an Associated Particle Imaging system to measure carbon in soil and other elements. Specifically, we present results from a pre-mixed soil sample containing pure sand (SiO$_2$) and \SI{4}{\percent} carbon by weight. Because the main isotopes of all those three elements emit characteristic high-energy gamma rays following inelastic neutron scattering, it is possible to measure their distribution with our instrument. A 3D resolution of several centimeters in all dimensions has been demonstrated.
\end{abstract}


\section{INTRODUCTION}
\label{sec:intro}
\begin{figure}[ht]
   \begin{center}
   \includegraphics[width=0.8\linewidth]{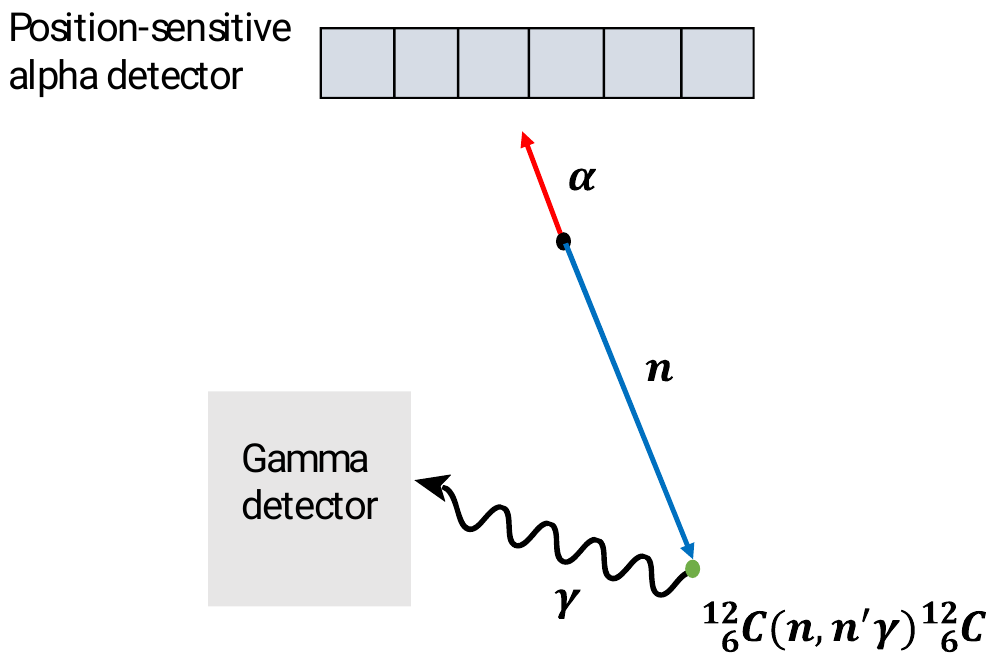}
   \end{center}
   \caption{\label{fig:API-diagram} Schematic illustration of API. A \SI{14.1}{\MeV} neutron and an alpha particle are created at a known location. The neutron undergoes inelastic scattering off an isotope in the target, for example a $^{12}$C nucleus, which emits a \SI{4.4}{\MeV} gamma ray. The location of the interaction can be determined in 3D from the time difference between the detection of the alpha particle and the gamma ray and from the $xy$-measurement of the position of the alpha particle in the detector.}
\end{figure}

Associated Particle Imaging (API) has been used in the past for detection of illicit drugs \cite{FONTANA2017279}, explosives \cite{CARASCO2008397}, and special nuclear material (SNM) \cite{JI2015105} among other applications. At Lawrence Berkeley National Laboratory (LBNL), we developed an API instrument for quantifying carbon distributions in soil. Non-destructive carbon-in-soil measurement methods are important for understanding and quantifying soil-based carbon sequestration techniques that offer sequestration potential on a large scale since soil is the largest storage pool of terrestrial carbon \cite{Sanderman2017-yz}. Monitoring carbon in soil also supports improvements in soil health and crop yield \cite{Huang2018-md}. Towards these goals, our API system employs state-of-the-art components, including a custom neutron generator equipped with a position sensitive alpha-particle detector, lanthanum bromide (LaBr$_3$) and sodium iodine (NaI) gamma-ray detectors, and a fast digital data acquisition (DAQ) system, that allows for high-spatial resolution of several centimeters in all three dimensions. The basic API method is depicted in Figure~\ref{fig:API-diagram}.

Neutrons are generated in deuterium(D) -- tritium(T) fusion reactions (Equation~\ref{eq:reactionDT}) at beam energies of up to \SI{100}{\keV}.  The two DT fusion reaction products, the alpha particle and the neutron, are emitted at \SI{180}{\degree} from each other in the center-of-mass reference frame. Therefore, by detecting the alpha particle in a position-sensitive detector, it is possible to calculate the direction of the associated neutron. The distance the neutron has traveled can be estimated from the recorded time difference between the detection of the alpha particle and the associated neutron-induced gamma ray. Taken together one can reconstruct a full 3D distribution of the irradiated volume.
\begin{equation}\label{eq:reactionDT}
    D + T \rightarrow \alpha\ (\SI{3.5}{\MeV}) + n\ (\SI{14.1}{\MeV})
\end{equation}

This technique offers an opportunity to non-destructively measure carbon distributions in heterogeneous soils, including changes with depth, and providing much larger representative samples than what is currently possible with standard core sampling techniques. Changes over time can be observed by repeatedly measuring the same location.

\section{SYSTEM COMPONENTS}
\label{sec:components}
The main components of the API system include the neutron generator, the alpha detector, two gamma detectors (LaBr$_3$ and NaI), and the DAQ system.

\subsection{Neutron generator}
\label{sec:neutron_generator}
\begin{figure}[ht]
   \begin{center}
   \includegraphics[width=\linewidth]{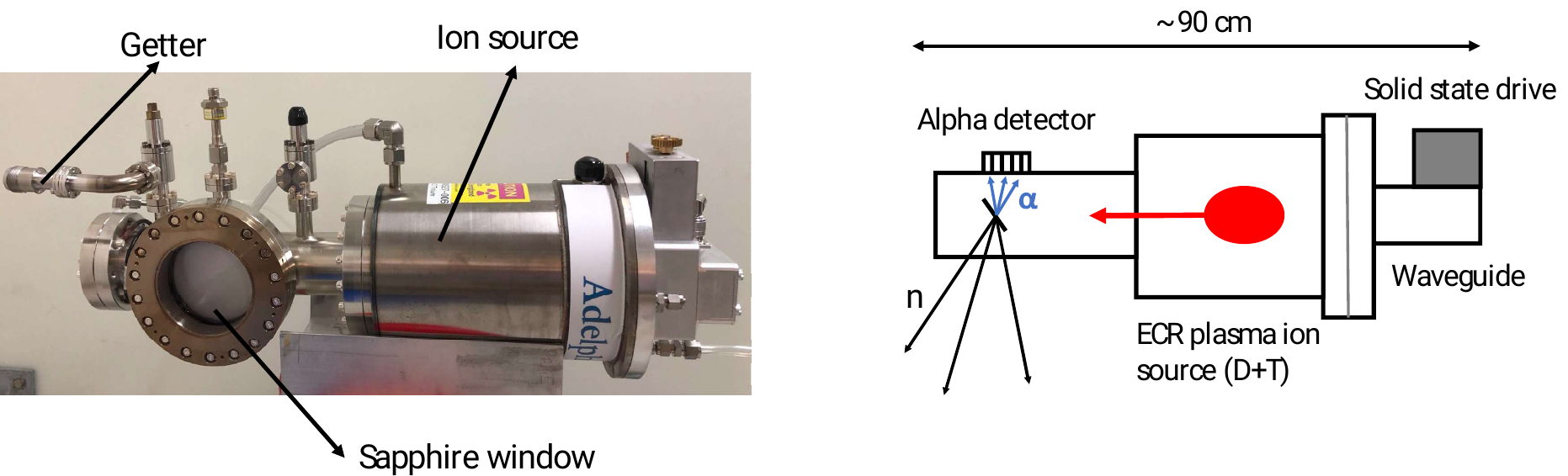}
   \end{center}
   \caption{\label{fig:ngen} Left: image of the DT108-API neutron generator tube and right: conceptual illustration of neutron generation and alpha detection.}
\end{figure}

The neutron generator was designed and built by Adelphi Inc\cite{Adelphi} based on their DT108-API generator. It is a sealed-type, compact neutron generator that consists of a microwave-driven ion source floated at a positive electrical potential, a high-voltage power supply with a maximum output of \SI[retain-explicit-plus]{+100}{\kilo\volt}, a hydrogen getter that controls the gas pressure inside the chamber, and a titanium target explosion-bonded to a copper backing at ground potential. A sapphire vacuum window located \SI{6}{\cm} from the target enables light transport from the scintillator detecting the alpha particle to the photomultiplier tube (PMT). The neutron generator together with a schematic drawing is shown in Figure~\ref{fig:ngen}. The neutron generator combines a high output of \SI{2e8}{neutrons\per\s} for high counting rates with a very small beam spot (\SI{2}{\mm} diameter) on target for excellent spatial resolution.

\subsection{Alpha detector}
The alpha detector consists of several components: an inorganic scintillator crystal, Yttrium Aluminum Perovskite (YAP), with dimension $\SI{50}{\mm}\times\SI{50}{\mm}\times\SI{1}{\mm}$ (Crytor \cite{Crytur}) which is placed approximately \SI{6}{\cm} from the beam target inside the neutron generator. The photons generated inside the YAP are transported through a \SI{3}{\mm} vacuum sapphire window (MPF Products \cite{MPF}) to the photocathode of a PMT (Hamamatsu \mbox{H13700-03} \cite{Hamamatsu}) located outside the vacuum chamber. An aluminum coating on the beam-facing side of the YAP acts as a mirror and doubles the amount of photons arriving at the PMT. The coating also blocks back-scattered deuterium and tritium ions from hitting the scintillator. The PMT is currently read out with a four-corner readout board enabling accurate measurement of the position (\SI{<1}{\mm} on the alpha-detector, corresponding to \SI{<5}{\cm} at \SI{60}{\cm} from the neutron source). For more details on the alpha detector please see our previous publication \cite{Unzueta2018-rk}. However, the current board design also limits the maximum allowable alpha detection rate to about \SI{2.5e5}{alphas\per\second} (corresponding to \SI{5e6}{neutrons\per\second} taking the solid angle of the alpha detector into account) due to longer RC delays of the measured signals and the fact that each alpha particle creates a signal on all four channels. Both facts resulting in pile-up issues at higher rates.  Future experiments will utilize a new board design to allow operation at higher rates.

\subsection{Lanthanum bromide and sodium iodide gamma detectors}
\label{sec:gamma-detectors}
The gamma detectors, both inorganic scintillators, were chosen because of their fast scintillation light decay time and relatively good energy resolution. The \SI{3}{inch} LaBr$_3$ detector (Saint Gobain B390S) excels at both of these requirements since its scintillation light has a \SI{25}{\nano\second} decay time, and its energy resolution at \SI{4.4}{\MeV} is approximately \SI{1}{\percent}. Additionally, we utilize a larger (\SI{5}{inch}) NaI detector (Alpha Spectra 20I20/5(9823)BN). However, LaBr$_3$ is significantly more expensive than the NaI scintillators. Scintillation light form NaI has a decay time of \SI{120}{\nano\second} and a resolution of \SI{3}{\percent} at \SI{4.4}{\MeV}. The NaI detector performance will be compared to that of the LaBr$_3$ detector, and it will be used to boost the count rate from deeper soil layers where we might be able to trade-off the timing precision for total measurement time.

\begin{figure}[!bth]
   \begin{center}
   \includegraphics[width=0.9\linewidth]{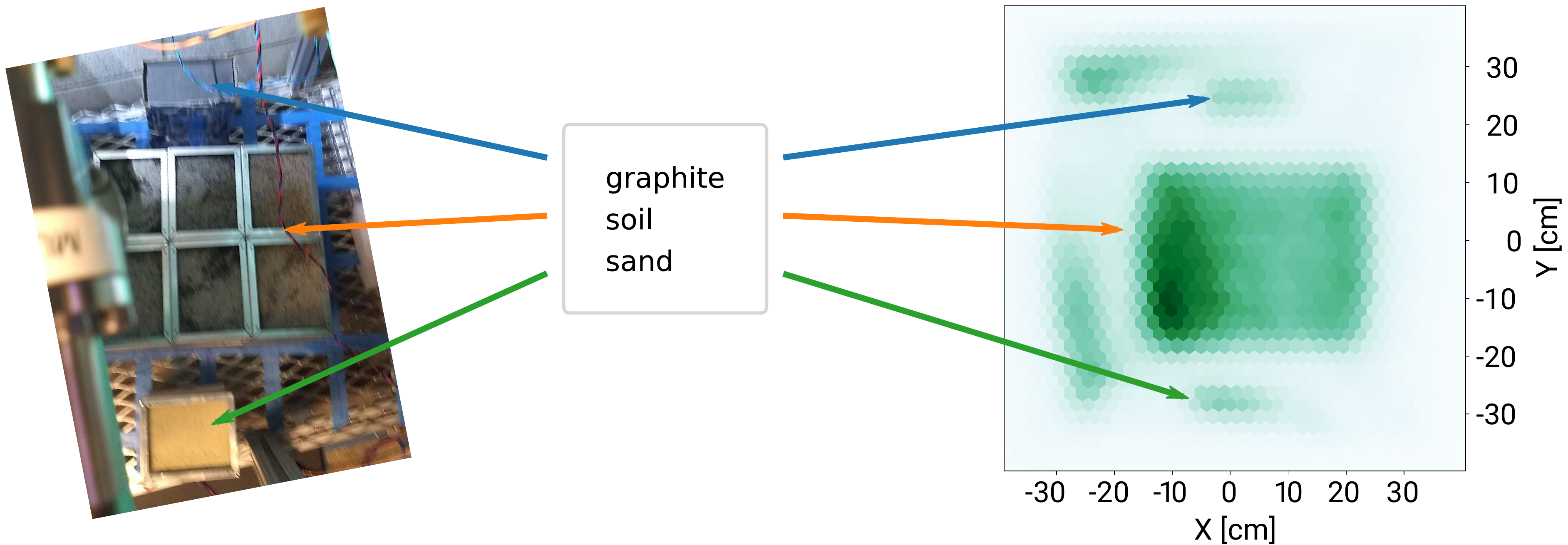}
   \end{center}
   \caption{\label{fig:xy-result}API measurements of soil, graphite, and sand samples. Left: experimental setup; right: $x$-$y$ distribution of detected events. The measurement also shows pickup of events from the wall and some neutron shielding on the left side.}
\end{figure}

\subsection{Data acquisition system}
For data acquisition and pre-processing, we use the PIXIE-16\cite{xia_manual-2018}, which is a 16-channel digital DAQ. It is capable of complex coincidence logic between multiple channels, implements an energy filter with on-the-fly pileup detection, as well as a constant fraction discrimination (CFD) algorithm for time-of-arrival detection. It consists of 16 analog-digital converters (\SI{500}{\MHz}) that feed into four field-programmable gate arrays (FPGAs) and a single digital signal processor (DSP) for the online calculation. In list mode, events consisting of time-stamped energies are buffered locally and then read out to a computer and written to disk.

The time of arrival is obtained with a CFD algorithm, which produces an output signal a fixed time after the leading edge of the pulse has reached a constant fraction of the pulse amplitude. This allows for the timing signal to be amplitude independent over a wide energy range \cite[p. 662]{knoll2010radiation}.
Measurements (in a forthcoming publication) using a \SI{1}{inch} thick dry sand sample where used to obtain the full width half maximum (FWHM) of the time difference between the alpha detection (YAP scintillator) and each of the gamma signals using a customized firmware for the PIXIE-16 system with optimized parameters for the CFD algorithm. Time resolutions of \SI{1.3}{\nano\second} for the LaBr$_3$ and \SI{1.6}{\nano\second} for the NaI detector were measured this way. The measurement still includes contributions to the FWHM from the horizontal extend of the sample as well as a component due to the measurement in the alpha detector (estimated to be on the same order as the LaBr$_3$ detector). Further optimization of the parameters used in the CFD algorithm could potentially improve the time resolution to approximately \SI{1}{\nano\second} for the YAP - LaBr$_3$ detector combination. The currently achieved FWHM of \SI{1.3}{\nano\second} corresponds to a depth resolution of \SI{7}{\cm}.

\begin{figure}[!tbh]
   \begin{center}
   \includegraphics[width=0.7\linewidth]{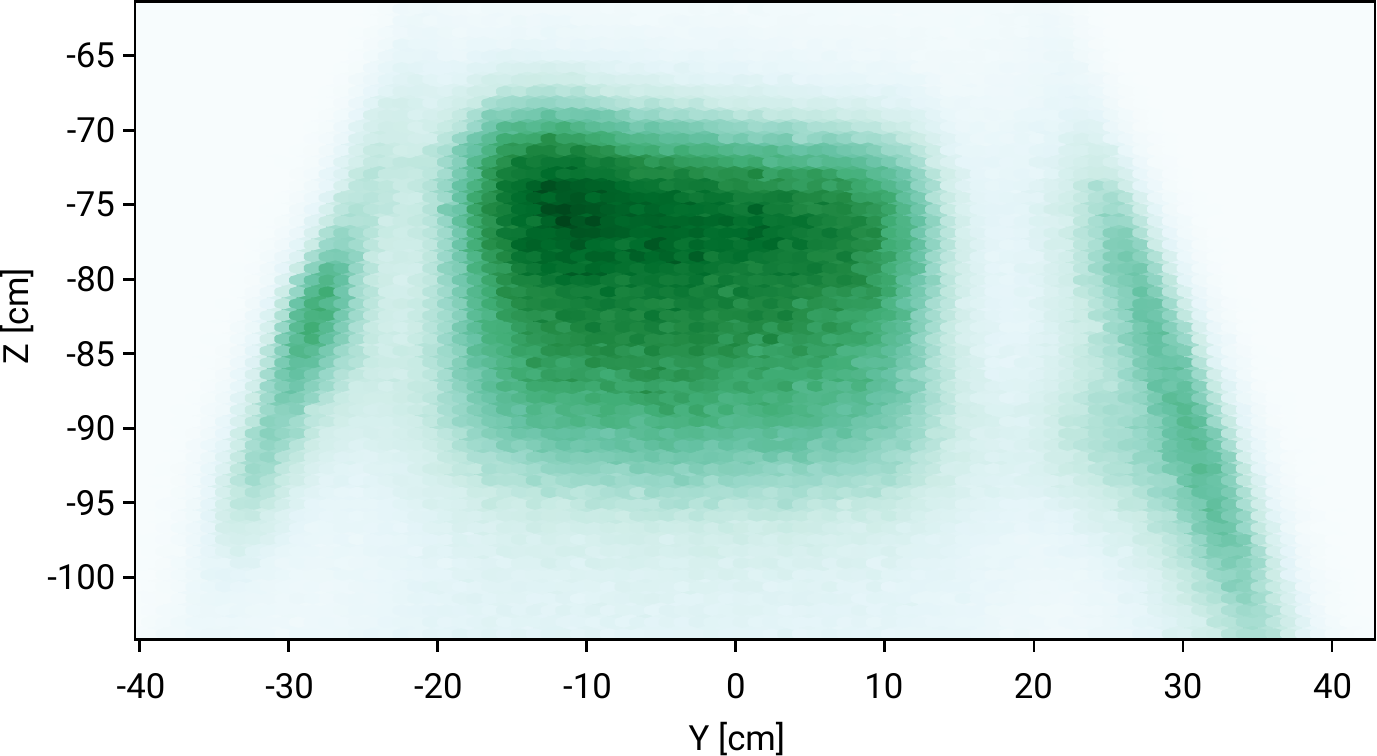}
   \end{center}
   \caption{\label{fig:yz-result}API measurements of the same setup as shown in Figure~\ref{fig:xy-result}, showing the $y$-$z$ projection. The neutron generator is located at a $z$-location of \SI{0}{\cm}. The API cone is clearly visible from some neutron shielding at the sides. Due to the contrast, the graphite and sand boxes are not clearly visible in this view.}
\end{figure}

\section{EXPERIMENTAL SETUP AND RESULTS}
\label{sec:experiment}
\begin{figure}[!bht]
   \begin{center}
   \includegraphics[width=0.8\linewidth]{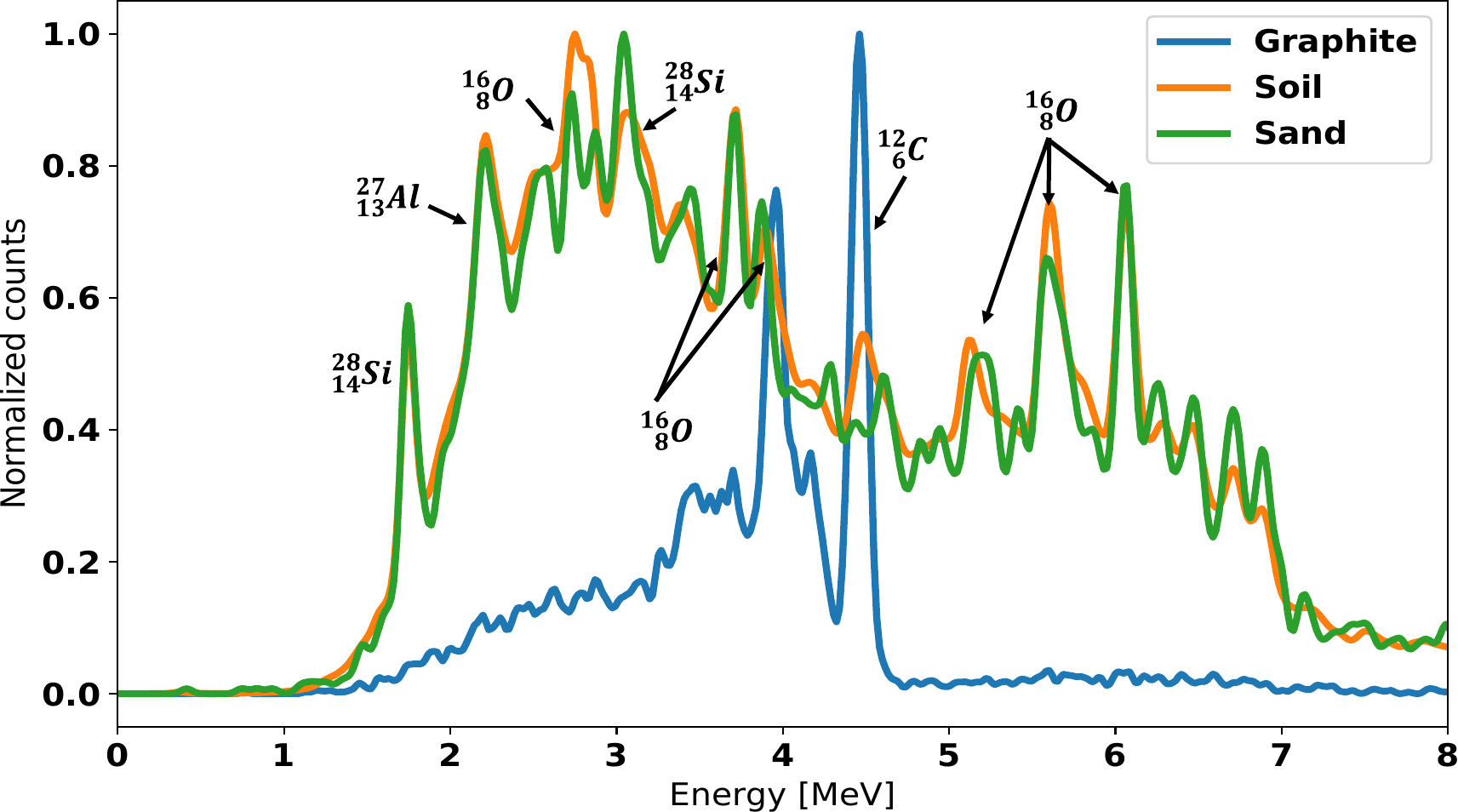}
   \end{center}
   \caption{\label{fig:spectra} LaBr$_3$ gamma spectra taken with of soil, sand, and graphite samples. See Figure~\ref{fig:xy-result} for details on the experimental setup. Gamma rays with energies \SI{<3}{\MeV} are suppressed due the CFD threshold being set too high in the PIXIE-16 DAQ. Also note that the resolution at \SI{4.4}{\MeV} is about \SI{3}{\percent}, which is higher than previously measured resolution for the same detector. This is most likely due to gain drift during the long measurement.}
\end{figure}

Sand (SiO$_2$) was mixed with \SI{4}{\percent} carbon by weight to simulate real soil and placed in several aluminum boxes (each \SI{3}{inch} $\times$ \SI{4}{inch} $\times$ \SI{5}{inch}). Well graded sand was sieved to remove fine pebbles then mixed with agricultural-grade worm castings, both oven-dried at \SI{175}{\celsius} to remove excess moisture. Worm castings have a relatively high carbon content (\SI{38.3}{\percent} C by weight as measured by loss on ignition method) and were thoroughly combined with sand to produce homogeneous, artificial soil mixtures of \SI{4}{\percent} carbon by weight. Boxes containing this mixture where stacked in a $2\times 3\times 3$ array centered below the neutron source with the top surface of the boxes located \SI{70}{\cm} from the neutron source, and irradiated at a neutron generator output of \SI{5e6}{neutrons\per\second} (acceleration voltage of \SI{50}{\kilo\volt}). Additionally, we placed a graphite brick of dimensions \SI{22}{cm} $\times$ \SI{13}{cm} $\times$ \SI{6.5}{cm} and another aluminum box filled with pure sand right next to the test soil to act as standards, as shown in Figure~\ref{fig:xy-result}. Data was acquired in coincidence mode for 9 hours (future improvement will greatly reduce this measurement time, see discussion in Section~\ref{sec:outlook}). Figure~\ref{fig:xy-result} shows the experimental setup and the reconstructed $x$-$y$ image and Figure~\ref{fig:yz-result} shows the $y$-$z$ projection of the same measurement. Spectra of the different regions in space according to the three different samples are shown in Figure~\ref{fig:spectra}.

Note that the API technique allows for the measurement of gamma-ray spectra from selected regions in space, greatly reducing counts from unwanted events that would show up in other measurement techniques. This has proven to be useful in other areas such as the one reported by Litvak et al. \cite{LITVAK201919} where spectroscopic analysis is rather difficult due to high background levels coming from structural materials. The gamma spectra in Figure~\ref{fig:spectra} shows several peaks which are the result of inelastic neutron scattering (INS) on the three main elements (Si, C, O), several transitions between excited states of each nuclei, and interactions of gamma rays within the detector crystal.
For instance, the graphite brick spectrum shows mostly the result of the transition of the \SI{4.4}{\MeV} excited state in $^{12}$C to its ground state. The peak at \SI{3.9}{\MeV} is the single escape peak due to pair production within the crystal. Also notice the interfering gamma line from $^{28}$Si at \SI{4.5}{\MeV} in the soil and sand spectra that needs to be taken into account for an accurate measurement of carbon distribution in soil. The unique ability to obtain gamma spectra from specific regions of space isolated from structural materials and unwanted background is clearly exemplified in the gamma spectrum of the graphite brick above. Table~\ref{tab:energies} shows the main gamma peaks of interest identified in our soil spectrum.

\begin{table}[htb]
    \centering
    \begin{tabular}{cc} \hline \hline
        Element & Energy [\si{\MeV}] \\ \hline
        $^{28}$Si & 1.78 \\
        $^{27}$Al & 2.21 \\
        $^{16}$O & 2.74 \\
        $^{28}$Si & 2.84 \\
        $^{16}$O & 3.68 \\
        $^{16}$O & 3.85 \\
        $^{12}$C & 4.44 \\
        $^{28}$Si & 4.497 \\
        $^{16}$O & 6.13 \\ \hline \hline
    \end{tabular}
    \caption{\label{tab:energies}Main inelastic gamma lines associated with the most common elements in soil. One important note is that the $^{28}$Si peak is close to the $^{12}$C peak which it hard to resolve, especially for the NaI detector.}
\end{table}

\section{CONCLUSION AND OUTLOOK}
\label{sec:outlook}
First experimental results and characterization of capabilities where obtained for an API system developed at LBNL. We showed that by performing 3D reconstruction of the location of the neutron scattering centers using a position sensitive alpha detector together with time-of-flight analysis, we can achieve a \SI{<5}{\cm} lateral resolution at \SI{60}{\cm} from the neutron source and a depth resolution of \SI{<7}{\cm}. Additionally, we demonstrated the ability to acquire energy spectra from specific regions of interest thereby greatly reducing the unwanted events resulting from gammas produced outside the volume of interest. All three detectors show the expected energy and time resolution.

We are currently developing a new alpha detector readout board that is capable of handling higher rates (approximately $\times$ 5). Further planned changes to the readout board will allow us to operate at the maximum neutron generator output of \SI{2e8}{neutrons\per\second}. These changes will allow us to cut the measurement time for the shown measurement down to less than 15 minutes. Additional detectors could be employed to reduce the measurement time even further. We are also in the process of developing the analysis software to extract accurate carbon density information from the gamma spectra and neutron flux measurements. Finally, we designed together with Adelphi Inc. a new neutron generator with several improvements that will have an impact on the overall resolution of the system, data quality, and portability.

\acknowledgments
The authors also would like to thank Takeshi Katayanagi for mechanical
support. The information, data, or work presented herein was funded by the Advanced Research Projects Agency-Energy (ARPA-E), U.S. Department of Energy, under Contract No. DEAC02-05CH11231.

\bibliography{main}

\begin{thebibliography}{10}

\bibitem{FONTANA2017279}
C.~L. Fontana, A.~Carnera, M.~Lunardon, F.~Pino, C.~Sada, F.~Soramel,
  L.~Stevanato, G.~Nebbia, C.~Carasco, B.~Perot, A.~Sardet, G.~Sannie,
  A.~Iovene, C.~Tintori, K.~Grodzicki, M.~Moszyński, P.~Sibczyński,
  L.~Swiderski, and S.~Moretto, ``Detection system of the first rapidly
  relocatable tagged neutron inspection system (rrtnis), developed in the
  framework of the european h2020 c-bord project,'' {\em Physics Procedia} {\bf
  90}, pp.~279 -- 284, 2017.
\newblock Conference on the Application of Accelerators in Research and
  Industry, CAARI 2016, 30 October – 4 November 2016, Ft. Worth, TX, USA.

\bibitem{CARASCO2008397}
C.~Carasco, B.~Perot, S.~Bernard, A.~Mariani, J.-L. Szabo, G.~Sannie, T.~Roll,
  V.~Valkovic, D.~Sudac, G.~Viesti, M.~Lunardon, C.~Bottosso, D.~Fabris,
  G.~Nebbia, S.~Pesente, S.~Moretto, A.~Zenoni, A.~Donzella, M.~Moszynski,
  M.~Gierlik, T.~Batsch, D.~Wolski, W.~Klamra, P.~L. Tourneur, M.~Lhuissier,
  A.~Colonna, C.~Tintori, P.~Peerani, V.~Sequeira, and M.~Salvato, ``In-field
  tests of the euritrack tagged neutron inspection system,'' {\em Nucl.
  Instrum. Methods Phys. Res. A} {\bf 588}(3), pp.~397 -- 405, 2008.

\bibitem{JI2015105}
Q.~Ji, B.~Ludewigt, J.~Wallig, W.~Waldron, and J.~Tinsley, ``Development of a
  time-tagged neutron source for snm detection,'' {\em Physics Procedia} {\bf
  66}, pp.~105 -- 110, 2015.
\newblock The 23rd International Conference on the Application of Accelerators
  in Research and Industry - CAARI 2014.

\bibitem{Sanderman2017-yz}
J.~Sanderman, T.~Hengl, and G.~J. Fiske, ``Soil carbon debt of 12,000 years of
  human land use,'' {\em Proc. Natl. Acad. Sci. U. S. A.} {\bf 114},
  pp.~9575--9580, Sept. 2017.

\bibitem{Huang2018-md}
J.~Huang, B.~Minasny, A.~B. McBratney, J.~Padarian, and J.~Triantafilis, ``The
  location- and scale- specific correlation between temperature and soil carbon
  sequestration across the globe,'' {\em Sci. Total Environ.} {\bf 615},
  pp.~540--548, Feb. 2018.

\bibitem{Adelphi}
{Adelphi Technology Inc.} http://adelphitech.com/, 2018.

\bibitem{Crytur}
{CRYTUR, spol. s r.o.} https://crytur.cz, 2018.

\bibitem{MPF}
{MPF Products}. https://mpfpi.com/, 2017.

\bibitem{Hamamatsu}
{Hamamatsu Photonics K.K.} https://www.hamamatsu.com, 2018.

\bibitem{Unzueta2018-rk}
M.~A. Unzueta, W.~Mixter, Z.~Croft, J.~Joseph, B.~Ludewigt, and A.~Persaud,
  ``Position sensitive alpha detector for an associate particle imaging
  system,'' {\em arXiv.org} , p.~1811.08591, Nov. 2018.

\bibitem{xia_manual-2018}
XIA, {\em Pixie-16 User Manual}, Aug. 2018.

\bibitem{knoll2010radiation}
G.~Knoll, {\em Radiation Detection and Measurement}, Wiley, 3rd~ed., 2010.

\bibitem{LITVAK201919}
M.~Litvak, Y.~Barmakov, S.~Belichenko, R.~Bestaev, E.~Bogolubov,
  A.~Gavrychenkov, A.~Kozyrev, I.~Mitrofanov, A.~Nosov, A.~Sanin, V.~Shvetsov,
  D.~Yurkov, and V.~Zverev, ``Associated particle imaging instrumentation for
  future planetary surface missions,'' {\em Nucl. Instrum. Methods Phys. Res.
  A} {\bf 922}, pp.~19 -- 27, 2019.

\end{thebibliography}
\bibliographystyle{spiebib}

\end{document}